\documentclass[a4paper,11pt]{article}
\pdfoutput=1 

\usepackage{jheppub} 

\usepackage[T1]{fontenc} 

\usepackage{slashed}
\usepackage{bm}

\newcommand{\bi}{\begin{itemize}}
\newcommand{\ei}{\end{itemize}}
\newcommand{\bracket}[2]{\bra{#1}\,#2\rangle} 
\newcommand{\bra}[1]{\langle\,#1\,|}          
\newcommand{\ket}[1]{|\,#1\,\rangle}          
\newcommand{\ud}{\mathrm{d}}

\newcommand{\LCp}{{\scriptscriptstyle +}}

\newcommand{\LCperp}{{\scriptscriptstyle \perp}}

\newcommand{\spara}{{\scriptscriptstyle\parallel}}

\newcommand{\f}[1]{\mbox{\boldmath$#1$}}

\newcommand{\bea}{\begin{eqnarray}}
\newcommand{\ea}{\end{eqnarray}}
\newcommand{\eea}{\end{eqnarray}}

\newcommand{\be}{\begin{equation}}
\newcommand{\ee}{\end{equation}}

\title{\boldmath Dynamically assisted Sauter-Schwinger effect -- 
non-perturbative versus perturbative aspects}

\author{G. Torgrimsson}
\author{C. Schneider}
\author{J. Oertel}
\author{R. Sch\"utzhold}

\affiliation{Fakult\"at f\"ur Physik, 
Universit\"at Duisburg-Essen, Lotharstra{\ss}e 1, Duisburg 47048, Germany}

\emailAdd{greger.torgrimsson@uni-due.de}
\emailAdd{christian.schneider@uni-due.de}
\emailAdd{johannes.oertel@uni-due.de}
\emailAdd{ralf.schuetzhold@uni-due.de} 

\abstract{The Sauter-Schwinger effect predicts the creation of 
electron-positron pairs out of the quantum vacuum by a strong and slowly 
varying electric field.
This effect can be dynamically assisted by an additional weaker time-dependent 
field, which may drastically enhance the pair-creation probability.
In previous studies, it has been found that the enhancement may crucially 
depend on the temporal shape of this weaker pulse, e.g., a Gaussian profile 
$\exp\{-(\omega t)^2\}$ or a Sauter pulse $1/\cosh^2(\omega t)$ 
behave quite differently.
In order to understand this difference, we 
make a perturbative expansion in 
terms of the weaker field while treating the strong electric field 
non-perturbatively. 
For a large class of profiles including the Sauter pulse, already the sum 
of the zeroth-order and the first-order amplitudes of this perturbative 
expansion yields good agreement.
For other cases, such as a Gaussian or sinusoidal profile, this is not true 
in general and higher orders can yield the dominant contribution -- 
where the dominant order depends on the chosen parameters.  
Our findings are confirmed by numerical simulations and help us to sort 
previous results into a bigger picture.}

\begin{document} 
\maketitle
\flushbottom

\section{Introduction}

The Sauter-Schwinger effect~\cite{Sauter:1931zz,Heisenberg:1935qt,Schwinger:1951nm} is a fundamental prediction of quantum field theory 
and corresponds to the creation of electron-positron pairs out of the quantum 
vacuum induced by a strong and slowly varying electric field via tunneling. 
For a constant field $E$, the pair-creation probability scales as 
($\hbar=c=1$) 
\bea
\label{Sauter-Schwinger}
P_{e^+e^-}
\propto
\exp\left\{-\pi\,\frac{m^2}{qE}\right\}
=
\exp\left\{-\pi\,\frac{E_{\rm crit}}{E}\right\}
\,,
\ea
where $m$ is the mass and $q$ the charge of the electron. 
The above expression does not admit a Taylor expansion in $E$ (nor $q$), 
which already indicates that this is a non-perturbative effect. 
Unfortunately, because the critical field strength $E_{\rm crit}=m^2/q\approx 1.3\times 10^{18}\text{V/m}$ 
is very large, this prediction has not been verified yet -- in contrast to pair production in the perturbative regime, which has been observed at SLAC~\cite{Bamber:1999zt}.

This motivates the search for ways to enhance the pair-creation probability 
$P_{e^+e^-}$, such as colliding multiple pulses~\cite{Bulanov:2010ei} or assisting the strong field with a high-energy photon~\cite{Dunne:2009gi}.
Another possibility is the dynamically assisted Sauter-Schwinger effect~\cite{Schutzhold:2008pz}, which has become an active research field, see e.g.~\cite{Orthaber:2011cm,Fey:2011if,Schneider:2014mla,Otto:2014ssa,Otto:2015gla,Linder:2015vta,Panferov:2015yda,Schneider:2016vrl}.   
To explain this phenomenon, let us imagine superimposing a slowly varying and 
strong field with a weaker time-dependent field, for example in the shape of 
two Sauter pulses~\cite{Schutzhold:2008pz}  
\bea
\label{double-Sauter}
\f{E}(t)=E\left[
\frac{1}{\cosh^2(\omega_{\rm slow}t)}
+
\frac{\varepsilon}{\cosh^2(\omega_{\rm fast}t)}
\right]\f{e}_z 
\,,
\ea
with $\omega_{\rm slow}\ll\omega_{\rm fast}\ll m$ and $E\ll E_{\rm crit}$ 
as well as $\varepsilon\ll1$. 
Here the Keldysh parameter for the strong and slow pulse 
$\gamma_{\rm slow}=m\omega_{\rm slow}/(qE)$ is supposed to be small $\gamma_{\rm slow}\ll1$ while the Keldysh parameter for the weaker and faster pulse 
$\gamma_{\rm fast}=m\omega_{\rm fast}/(qE\varepsilon)$ is large 
$\gamma_{\rm fast}\gg1$. 
To understand the combined impact of the two pulses (dynamical assistance) it is convenient to consider the combined Keldysh parameter~\cite{Schutzhold:2008pz} 
\bea
\label{Keldysh}
\gamma=\frac{m\omega_{\rm fast}}{qE} \;.
\ea
For $\gamma$ below the threshold $\gamma_{\rm crit}=\pi/2$ the weak field has negligible impact and we obtain basically the same result as in~\eqref{Sauter-Schwinger}. Above threshold $\gamma>\gamma_{\rm crit}$, however, the weak field can significantly enhance the pair creation probability $P_{e^+e^-}$. The existence of such a threshold is characteristic for dynamical assistance.	

However, the value of the threshold can depend strongly on the shape of the weaker pulse~\cite{Linder:2015vta}. Indeed, for a Gaussian profile $\sim\exp\{-(\omega_{\rm fast}t)^2\}$ the threshold scales as $\gamma_{\rm crit}\sim\sqrt{|\ln\varepsilon|}$~\cite{Linder:2015vta}. So, while a Sauter and a Gaussian pulse are visually almost indistinguishable they lead to very different physics. In order to understand this difference we employ a perturbative expansion in terms of the weaker field. 

This paper is organized as follows. In section~\ref{Perturbative Approach} we give a brief description of our perturbative approach and present our final analytical results for the first orders and compare them with numerical simulations. In section~\ref{Higher orders section} we study higher orders and show that these can in some cases be larger than the first orders. In section~\ref{Derivation first order spectrum} we explain in more details how to derive the zeroth and first order amplitudes, and in section~\ref{HigherOrderDerivation} we show how to derive higher orders using results from the worldline formalism. In section~\ref{numerics} we outline the numerical method used to check our analytical approximations. We conclude in~\ref{Conclusions}. In appendix~\ref{First order from the polarization tensor} we explain how our first order results for the total probability can be recovered using the polarization tensor, and in appendix~\ref{Higher orders from worldline instantons} we rederive our higher order results using worldline instantons, which also allows us to generalize to spatially inhomogeneous fields.

\section{Perturbative Approach}\label{Perturbative Approach}

We consider spatially homogeneous but time-dependent electric fields of the following general form 
\bea
\label{general-form}
\f{E}(t)=E\left[f_0(t)+\varepsilon f_1(t)\right]\f{e}_z  \;,
\ea
where $f_0(t)$ is the strong and slow field and $f_1(t)$ denotes the weaker and faster pulse with $\varepsilon\ll1$. The idea now is to calculate the pair production probability via a Taylor expansion in $\varepsilon$. There are different methods for doing so, for example the worldline formalism (see section~\ref{HigherOrderDerivation}, appendix~\ref{Higher orders from worldline instantons} and~\cite{NonperturbativePerturbative2}). Here we start with a more conventional method based on WKB and standard time-dependent perturbation theory in the Furry picture, see section~\ref{Derivation first order spectrum} for more details. 

To this end we employ the interaction picture where the 
strong field $Ef_0(t)$ gives the $\hat H_0$-dynamics while the weaker field 
$E\varepsilon f_1(t)$ enters via the interaction Hamiltonian 
\bea
\label{interaction-Hamiltonian}
\hat H_{\rm int}(t)
=
q\int\ud^3r\,
\hat{\bar\Psi}(t,\f{r})
\gamma^\mu A_{\mu}^{\rm fast}(t) 
\hat\Psi(t,\f{r}) \;.
\ea
Accordingly the fermionic field operator $\hat\Psi(t,\f{r})$ solves the Dirac 
equation in the presence of the strong field $Ef_0(t)$ while 
$\dot A_{\mu}^{\rm fast}(t)=[0,0,0,E\varepsilon f_1(t)]$ 
encodes the weaker field (both in temporal gauge).  
Due to the spatial homogeneity, the $\ud^3r$-integral yields momentum conservation, 
i.e., electrons and positrons are created in pairs of opposite 
(canonical) momenta of equal magnitude $\f{p}_{e^+}=-\f{p}_{e^-}$. After standard manipulations (see section~\ref{Derivation first order spectrum}) the pair creation probability is given by
\bea
P_{e^+e^-}=V_3\int\frac{\ud^3p}{(2\pi)^3}\,
\left|
\mathfrak{A}_0(\f{p})+\varepsilon\,\mathfrak{A}_1(\f{p})
+\varepsilon^2\,\mathfrak{A}_2(\f{p})+\dots
\right|^2 \;,
\ea
where $\f{p}=\f{p}_{e^+}=-\f{p}_{e^-}$ and $V_3$ denotes the three-dimensional volume. 
The zeroth order amplitude $\mathfrak{A}_0(\f{p})$ is fully determined by the
$\hat H_0$-dynamics, i.e., the strong field alone, and the higher order amplitudes come from the weaker pulse. 
For a strong field in the shape of a Sauter pulse, $\mathfrak{A}_0(\f{p})$ can be obtained analytically from the exact solution of the Dirac equation in terms of hypergeometric functions~\cite{NarozhnyiNikishovSimplest,Nikishov1985,Hebenstreit:2011pm,Hebenstreit:2010vz}.
The first order $\mathfrak{A}_1(\f{p})$ can be obtained from standard time-dependent perturbation 
theory w.r.t.\ the interaction Hamiltonian~\eqref{interaction-Hamiltonian}.
Rewriting the time integral $\int dt\,\hat H_{\rm int}(t)$ as a frequency integral, 
we find 
\bea
\label{first-order-amplitude}
\mathfrak{A}_1(\f{p})=
\int\frac{\ud\omega}{2\pi}\,\tilde f_1(\omega)\,W_1(\omega,\f{p}) \;,
\ea
where $\tilde f_1(\omega)$ is the Fourier transform of the weaker pulse. In order to calculate the remaining matrix elements $W_1(\omega,\f{p})$ we treat the slow field with a WKB approach, c.f.~\cite{Wollert:2015oea,Torgrimsson:2016ant,DiPiazza:2016maj}. Due to $\omega_{\rm slow}\ll\omega_{\rm fast}$ the time during which the weaker field is operative is very short and hence we can approximate the slow field by a constant one. 
Then the remaining matrix element $W_1(\omega,\f{p})$ behaves as
\be\label{exponentinW1}
W_1(\omega,\f{p})\sim\exp\Big\{-\frac{m_\LCperp^2}{qE}\Big(\frac{\pi}{2}+i\phi\Big[\frac{i\omega}{2m_\LCperp}\Big]\Big)-i\frac{\omega p_\spara}{qE}+i\frac{m_\LCperp^2}{qE}\phi\Big[\frac{p_\spara}{m_\LCperp}\Big]\Big\}
\ee
with $m^2_\LCperp=m^2+\f{p}_\LCperp^2=m^2+p_x^2+p_y^2$, $p_\spara=p_z$, and the function we recognize from~\cite{Schutzhold:2008pz,Fey:2011if,Linder:2015vta}
\bea
\label{phi-function}
\phi(z)=z\sqrt{1+z^2}+{\rm arsinh}(z) \;,
\ea
which comes from the classical action of a particle in a constant electric field, see eq.~5 in~\cite{NikishovConstant}. 
At $\f{p}=0$ the exponent simplifies to
\be\label{firstorderW}
W_1(\omega,0)\sim\exp\Big\{-\frac{m^2}{qE}\Big(\frac{\pi}{2}+i\phi\Big[\frac{i\omega}{2m}\Big]\Big)\Big\}=\exp\Big\{-\frac{m^2}{qE}\Big(\frac{\pi}{2}-\frac{\omega}{2m}\sqrt{1-\Big[\frac{\omega}{2m}\Big]^2}-\arcsin\frac{\omega}{2m}\Big)\Big\} \;.
\ee
At $\omega=0$ we recover half the Schwinger exponent in~\eqref{Sauter-Schwinger} (before squaring the amplitude). As $\omega$ increases the exponential increases, and at $\omega=2m$ the amplitude stops being exponentially suppressed as expected. The Fourier transform in~\eqref{first-order-amplitude}, on the other hand, decreases as $\omega$ increases, so the $\omega$-integral in~\eqref{first-order-amplitude} is typically dominated by some $\omega_{\rm dom}$ that depends on the field shape.

\subsection{Sauter Pulse}

\begin{figure}
\centering
\includegraphics[width=.49\textwidth]{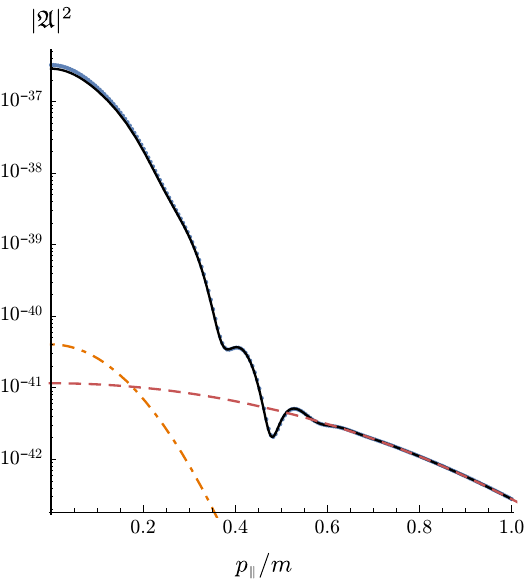}
\includegraphics[width=.49\textwidth]{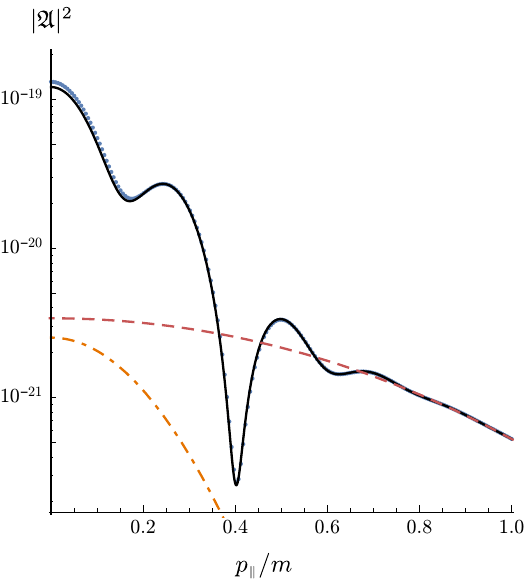}
\caption{Logarithmic plots of the pair production probability as a function of the longitudinal momentum $p_\spara/m$, with $p_\LCperp=0$, for the double Sauter pulse in~\eqref{double-Sauter} with the parameters $E=0.033E_{\rm crit}$ and $\gamma=2.2$ for the left plot, and $E=0.066E_{\rm crit}$ and $\gamma=2.43$ for the right plot, and $\gamma_{\rm slow}=0.2$ and $\varepsilon=10^{-3}$ in both cases. The solid black curves show the first order result $|\mathfrak{A}_0(\f{p})+\varepsilon\,\mathfrak{A}_1(\f{p})|^2$, where $\mathfrak{A}_0\approx-\beta^*$ is obtained from~\eqref{alphabetaSauter} and $\mathfrak{A}_1$ is obtained from~\eqref{M1SauterSauter} (which gives a slightly better approximation than~\eqref{M1result}). 
The blue dots correspond to the numerical solution of the Riccati equation, see section~\ref{numerics}. The red dashed curves give the result for the strong field alone, and the orange dot-dashed curves correspond to weak field alone. The value of $\gamma$ is chosen such that the dynamically assisted probability is much larger than the ones with either the strong or the weak field alone (the weak/strong field drops out for low/high $\gamma$). 
The spectrum is symmetric around $p_\spara=0$.}
\label{SauterSpec033-fig}
\end{figure}

In order to evaluate~\eqref{first-order-amplitude} we have to specify the shape of the weak field.
For a Sauter pulse $f_1(t)=1/\cosh^2(\omega_{\rm fast}t)$ 
as in Eq.~\eqref{double-Sauter}, the Fourier transform reads 
\bea\label{FourierSauterExact}
\tilde f_1(\omega)=\frac{\pi}{\omega_{\rm fast}^2}\,
\frac{\omega}{\sinh(\pi\omega/[2\omega_{\rm fast}])} \;.
\ea
For large $\omega\gg\omega_{\rm fast}$ it behaves as 
\bea\label{FourierSauterLargew}
\tilde f_1(\omega)
\sim
\exp\left\{-\frac{\pi}{2}\,\frac{\omega}{\omega_{\rm fast}}
\right\} \;.
\ea
This exponential decay is a general feature of a large class of pulses (such as a Lorentzian profile), which have poles at imaginary 
times $t_*=\pm i\tau_*$.  
To cover the general case we 
introduce the associated Keldysh parameter (see also~\cite{Torgrimsson:2016ant})
\bea
\gamma_*=\frac{m}{qE\tau_*} \;.
\ea
For a Sauter pulse we have $\tau_*=\pi/(2\omega_{\rm fast})$ and $\gamma_*=2\gamma/\pi$. 
Now we can estimate the $\omega$-integral in \eqref{first-order-amplitude} 
via the saddle-point method where $m^2/(qE)\gg1$ plays the role of the large parameter.
The saddle point gives us the dominant frequency $\omega_{\rm dom}$, i.e., the spectral content of the weaker field that gives the dominant contribution to the probability.
For $\gamma_*>1$, i.e., above threshold, we find a dominant frequency of
\bea\label{omegadomSauter1}
\omega_{\rm dom}
=
2\sqrt{m_\LCperp^2+\Big(p_\spara-\frac{im}{\gamma_*}\Big)^2}
=
2m_\perp\sqrt{1-\Big(
\frac{1}{\gamma_\LCperp}
+
\frac{ip_\spara}{m_\LCperp}
\Big)^2}
 \;,
\ea
where $\gamma_\LCperp=\gamma_* m_\perp/m$. For $p_\spara=0$,
the dominant frequency starts at zero at the threshold $\gamma_\LCperp=1$
with infinite slope and approaches $2m_\perp$ for large $\gamma_\LCperp$.
Inserting the saddle point at $\omega=\omega_{\rm dom}$ we find (for fixed spin, see section~\ref{FirstOrderDerivation})
\be\label{frakA1result}
\mathfrak{A}_1(\f{p})=4\pi\frac{qE}{\omega_{\rm fast}^2}\frac{m_\LCperp}
{\omega_{\rm dom}}
\exp\Big\{\!-\frac{im_\LCperp^2}{qE}\Big(\phi\Big[\frac{p_{\scriptscriptstyle\parallel}}{m_\LCperp}-\frac{i}{\gamma_\LCperp}\Big]-\phi\Big[\frac{p_{\scriptscriptstyle\parallel}}{m_\LCperp}\Big]\Big)\Big\} \;,
\ee 
with $\phi$ from Eq.~\eqref{phi-function}.
Together with the zeroth order this result already\footnote{It is clear from the agreement with numerical results that the cross term $\text{Re }\varepsilon^2\mathfrak{A}_0^*\mathfrak{A}_2$ is not important for these fields. We will study this in~\cite{NonperturbativePerturbative2} using the worldline formalism, see~\cite{Dumlu:2011cc}.} agrees well with our numerical simulations, see figure~\ref{SauterSpec033-fig}. In~\cite{Orthaber:2011cm} the momentum spectrum (for different parameters) was obtained numerically using the quantum kinetic formalism, and in~\cite{Fey:2011if} a WKB approximation was used to obtain exponentials that could qualitatively explain the spectrum in~\cite{Orthaber:2011cm}. However, the pre-exponential factors were only obtained by approximating them to be constants and matching these constants with the numerical results~\cite{Orthaber:2011cm}. Here we provide explicit analytical expressions for the prefactors, including their momentum dependence.

In the region of strong assistance, 
$\varepsilon\mathfrak{A}_1(\f{p})\gg\mathfrak{A}_0(\f{p})$, we can neglect the zeroth order contribution.
Then using that the momentum integral is peaked at $\f{p}=0$ (summing over spins gives a factor of $2$) we can approximate 
\be\label{P2-Sauter-integrated}
\begin{split}
P_{e^+e^-}&
\approx
V_3\!\int\!\frac{\ud^3p}{(2\pi)^3}\,
\left|\varepsilon\,\mathfrak{A}_1(\f{p})\right|^2 \\
&
\approx
\varepsilon^2\frac{V_3m^3}{\pi^2\gamma^2}\,
\sqrt{\frac{2\pi m^2}{qE}}\;
\frac{
\mbox{$\displaystyle
\exp\left\{
\!-\frac{2m^2}{qE}
\left(\frac{\sqrt{\gamma_*^2-1}}{\gamma_*^2}+\arcsin\frac{1}{\gamma_*}\right)
\right\}$}
}{
\mbox{$\displaystyle
(\gamma_*^2-1)^\frac{3}{4}\arcsin\frac{1}{\gamma_*}
$}} 
\;.
\end{split}
\ee
Note that this expression is perturbative in the weaker field $\varepsilon$, but non-perturbative in the strong field $E$. Importantly, we recover the exponent in~\cite{Schutzhold:2008pz} where both fields were treated nonperturbatively. Note also that the exponential expressed in terms of $\gamma_*$ has the same form for all Sauter-like fields, unlike the prefactor. 
Clearly, the prefactor in \eqref{P2-Sauter-integrated} breaks down as 
$\gamma_*$ approaches unity, i.e., the threshold, where the saddle point method 
for calculating the above integrals fails. 
Thus, $\gamma_*$ should lie sufficiently above unity, which is consistent with
our earlier assumptions:
in order to simplify the Fourier transform of the weak field as 
in~\eqref{FourierSauterLargew} we have assumed that 
$\omega/\omega_{\rm fast}\gg1$ and hence also 
$\omega_{\rm dom}\gg\omega_{\rm fast}$. 
Comparison with~\eqref{omegadomSauter1} shows that $\gamma_*$ should not be 
too close to unity, i.e., $m^2\sqrt{\gamma_*^2-1}\gg qE$, which is satisfied in the region with maximum enhancement. 

To check \eqref{P2-Sauter-integrated} we consider $\gamma_*\to\infty$, where the strong field drops out (i.e. $E$ only enters via the field strength of the weak field $E\varepsilon$),
\be\label{HighGammaSauter}
\lim\limits_{\gamma_*\to\infty}\eqref{P2-Sauter-integrated}=\frac{V_3m^3}{\pi}\sqrt{\frac{m}{\omega_{\rm fast}}}\left(\frac{qE\varepsilon}{m\omega_{\rm fast}}\right)^2e^{-2\pi m/\omega_{\rm fast}} \;,
\ee
which coincides with the exact result for a single Sauter pulse~\cite{NarozhnyiNikishovSimplest,Nikishov1985,Dunne:1998ni}, see eq.~2.3 in~\cite{Dunne:2004nc}, after a Taylor expansion to second order in $\varepsilon$. We can also verify~\eqref{HighGammaSauter} using eq.~38 in~\cite{Popov72}, which gives the probability to second order in the field strength of a single electric field (the weak field without the strong field in our case) in terms of an integral over the square of the Fourier transform. The exponential suppression in~\eqref{HighGammaSauter} in $\omega_{\rm fast}$ comes directly from the exponential suppression of the Fourier transform~\eqref{FourierSauterLargew} at $\omega=2m$. 

\begin{figure}
\centering
\includegraphics[width=.49\textwidth]{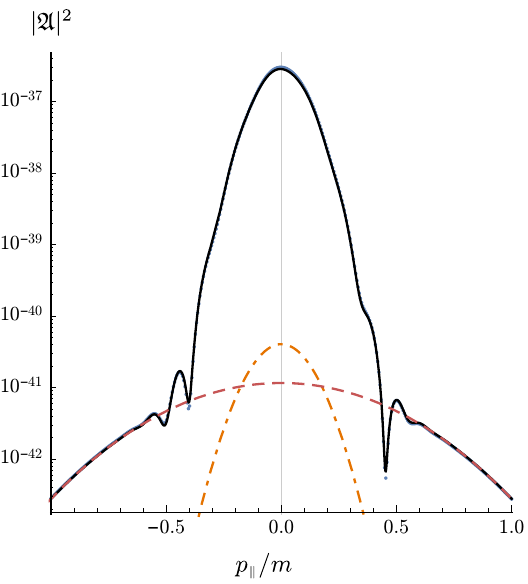}
\includegraphics[width=.49\textwidth]{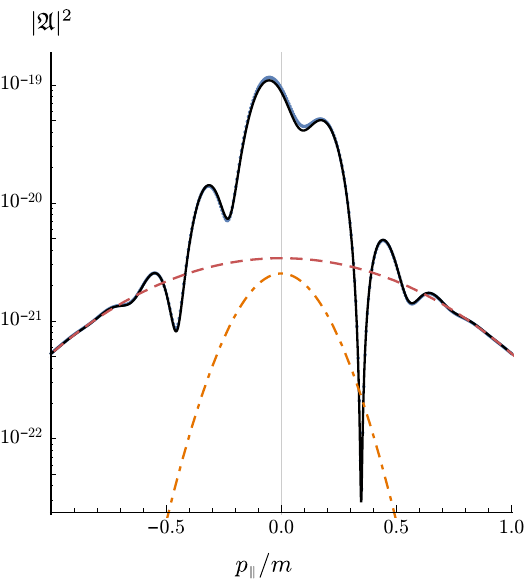}
\caption{Logarithmic plots of the pair production probability as a function of the longitudinal momentum $p_\spara/m$, with $p_\LCperp=0$.
The strong field is a Sauter pulse and the weak field is given by~\eqref{tanhsech-def}. The parameters are $E=0.033E_{\rm crit}$ and $\gamma=2.2$ for the left plot, $E=0.066E_{\rm crit}$ and $\gamma=2.43$ for the right plot, and $\gamma_{\rm slow}=0.2$ and $\varepsilon=10^{-3}$ in both cases. The curves and dots are obtained as described in Fig.~\ref{SauterSpec033-fig}, where $\mathfrak{A}_0\approx-\beta^*$ is obtained from~\eqref{alphabetaSauter} and $\mathfrak{A}_1$ is obtained by multiplying~\eqref{M1SauterSauter} with a factor of $-i$. Our first order result clearly agrees well with the exact numerical result.}
\label{TanhSech033-fig}
\end{figure}

As mentioned above, having obtained $\mathfrak{A}_1$ for a Sauter pulse, it is now easy to obtain $\mathfrak{A}_1$ for other weak fields with Fourier transforms with exponential decay. For example, consider a weak field given by
\be\label{tanhsech-def}
f_1(t)=-\frac{\sinh(\omega_{\rm fast}t)}{\cosh^2(\omega_{\rm fast}t)} \qquad\leadsto\qquad \tilde{f}_1(\omega)=-i\frac{\pi}{\omega_{\rm fast}^2}\frac{\omega}{\cosh(\pi\omega/[2\omega_{\rm fast}])} \;.
\ee
As a function of $t$, this field looks very different from a Sauter pulse. However, it has a similar Fourier transform. In fact, in the region $\omega\gg\omega_{\rm fast}$, the Fourier transform of this field is simply obtained by multiplying the Fourier transform of a Sauter field by a factor of $-i$. This means that the first order amplitude for this field is simply given by $\mathfrak{A}_1=-i\mathfrak{A}_1^\text{Sauter}$. This factor of $i$ leads to a quite different momentum spectrum because of interference, see figure~\ref{TanhSech033-fig}, but drops out in the region where $\mathfrak{A}_1$ is much larger than $\mathfrak{A}_0$.
%

\subsection{Gaussian Pulse}

As mentioned before, a Gaussian pulse shows qualitatively different behavior. So, as our
next example, we consider a Gaussian profile $f_1(t)=\exp\{-(\omega_{\rm 
fast}t)^2\}$. Then the
Fourier transform is also Gaussian 
\bea\label{GaussFourier}
\tilde f_1(\omega)=\frac{\sqrt{\pi}}{\omega_{\rm fast}}\,
\exp\left\{-\frac{\omega^2}{4\omega_{\rm fast}^2}\right\} \;.
\ea
For simplicity, we consider the peak of the momentum integral at $\f{p}=0$.
After the same steps as in the previous section, the
saddle-point method for the $\omega$-integral~\eqref{first-order-amplitude} yields
\be\label{domwGauss1}
\omega_{\rm dom}=\frac{2m}
{\sqrt{1+(qE/\omega_{\rm fast}^2)^2}} \;.
\ee
In contrast to the Sauter case considered in the previous section, here there is
no threshold for $\omega_{\rm dom}$, which starts at zero for $\omega_{\rm fast}=0$, 
first behaves quadratically
$\omega_{\rm dom}\approx2m\omega_{\rm fast}^2/(qE)$
for small $\omega_{\rm fast}$
and finally approaches $2m$ for large $\omega_{\rm fast}$. This
qualitative difference between a Sauter pulse and a Gaussian profile stems from the different Fourier transforms, especially at large $\omega$.
Now the saddle-point approximation for the momentum integral gives 
\bea\label{firstorderGaussian}
P_{e^+e^-}
\approx
\frac{V_3m^3}{32\sqrt{2\pi}}\left[\frac{qE\varepsilon}{m^2}\right]^2\frac{qE}{m\omega_{\rm fast}}\frac{\left(1+\left[\frac{qE}{\omega_{\rm fast}^2}\right]^2\right)^\frac{3}{2}}{\text{arctan}\frac{qE}{\omega_{\rm fast}^2}}\exp\Big\{\!-\frac{2m^2}{qE}\text{arctan}\frac{qE}{\omega_{\rm fast}^2}\Big\} \;.
\ea
%
For small $\omega_{\rm fast}$, $\arctan(qE/\omega_{\rm fast}^2)$ approaches $\pi/2$ and we recover the exponent in~\eqref{Sauter-Schwinger}.
If we now Taylor expand $\arctan(qE/\omega_{\rm fast}^2)$ for small $\omega_{\rm fast}^2\ll qE$, the exponent can be approximated by
\be
\exp\left\{
-\frac{2m^2}{qE}\,\arctan\frac{qE}{\omega^2_{\rm fast}}
\right\}\approx\exp\left\{
-\frac{\pi m^2}{qE}+2\gamma^2
\right\} \;.
\ee
This first order exponent is thus larger than the zeroth order exponent~\eqref{Sauter-Schwinger}, but the above contribution~\eqref{firstorderGaussian} also contains a factor of $\varepsilon^2$ in the prefactor. Demanding that this contribution be larger than the zeroth order term~\eqref{Sauter-Schwinger} gives us a threshold condition 
\be
\gamma\gtrsim\sqrt{|\ln\varepsilon|} \;,
\ee
which agrees with the threshold $\gamma_{\rm crit}\sim\sqrt{|\ln\varepsilon|}$ found in~\cite{Linder:2015vta} by treating both fields nonperturbatively. 

As in the Sauter case, we may consider the limit of large frequencies 
$\omega_{\rm fast}^2\gg qE$.
In this limit, the arctan function behaves linearly in $qE/\omega_{\rm fast}^2$
and thus the strong field drops out.
As a result, the exponential suppression,
\be\label{HighGammaGauss}
P_{e^+e^-}\approx\frac{V_3m^3}{32\sqrt{2\pi}}\left[\frac{qE\varepsilon}{m^2}\right]^2\frac{\omega_{\rm fast}}{m}\exp\left\{\!-\frac{2m^2}{\omega_{\rm fast}^2}\right\} \;,
\ee
is directly obtained from the exponential suppression of the Fourier transform~\eqref{GaussFourier} at $\omega=2m$. As for the Sauter case, \eqref{HighGammaGauss} agrees with eq.~38 in~\cite{Popov72}.   

\begin{figure}
\centering
\includegraphics[width=.49\textwidth]{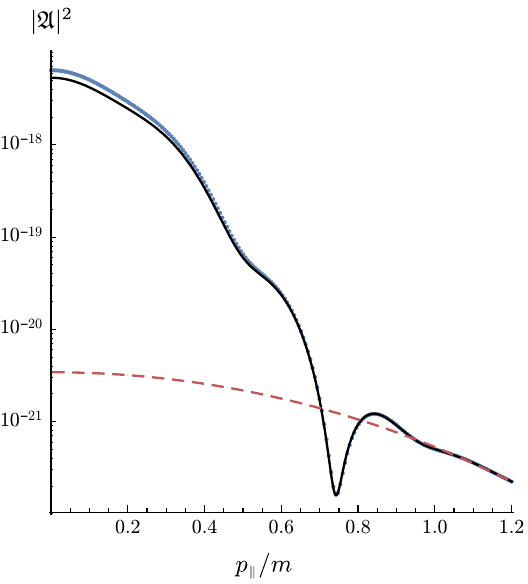}
\includegraphics[width=.49\textwidth]{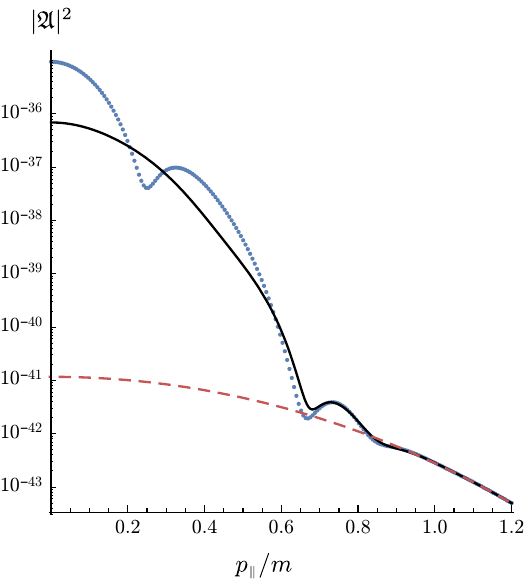}
\caption{Logarithmic plots of the pair production probability as a function of the longitudinal momentum $p_\spara/m$, with $p_\LCperp=0$.
The strong field is a Sauter pulse and the weak field is a Gaussian pulse~\eqref{GaussFourier}. The parameters are $E=0.066E_{\rm crit}$ and $\gamma=3.9$ for the left plot, $E=0.033E_{\rm crit}$ and $\gamma=3.8$ for the right plot, and $\gamma_{\rm slow}=0.2$ and $\varepsilon=10^{-3}$ in both cases. The curves are obtained as described in Fig.~\ref{SauterSpec033-fig}, where $\mathfrak{A}_0\approx-\beta^*$ is obtained from~\eqref{alphabetaSauter} and $\mathfrak{A}_1$ from~\eqref{M1DefGauss}. In the left plot we find a rather good agreement with the first order results. However, in the right plot there clearly are important contributions that are not captured by the first order.}
\label{GaussSpectra-fig}
\end{figure}

It is also possible to derive an analytic expression for the momentum spectrum for a Gaussian weak field, see~\eqref{M1DefGauss}. However, by plotting this first order spectrum and comparing it with the exact numerical spectrum, we find an interesting difference, see figure~\ref{GaussSpectra-fig}: As in the Sauter case, we consider two different field strengths, $E=0.066E_{\rm crit}$ and $E=0.033E_{\rm crit}$. For $E=0.066E_{\rm crit}$ the first order gives a good agreement with the numerical result, but for $E=0.033E_{\rm crit}$ the exact numerical spectrum is qualitatively different from the first order prediction. We show in the next section that this difference is due to higher order contributions becoming important for the parameters chosen in the second case.        
%

\section{Higher orders}\label{Higher orders section}

As we have seen above, the difference between a Sauter and a Gaussian pulse is already visible at first order. However, this difference is even more pronounced at higher orders. For deriving the higher orders in $\varepsilon$ we use the worldline formalism, see~\cite{Schubert:2001he} for a review.
This approach directly gives the pair production probability (instead of the amplitude), which is then Taylor expanded via 
\be\label{Pseriesdef}
P_{e^+ e^-}=P_0+\varepsilon P_1+\varepsilon^2 P_2+\dots
\ee 
Without a background field the Furry theorem says that all odd orders should vanish ($P_1=P_3=\dots=0$). With a background field, however, this is in general not true\footnote{Of course, the sum of the even orders is always larger than the sum of the odd orders, since the total probability~\eqref{Pseriesdef} remains positive even if we change the sign of $\varepsilon$.}~\cite{BialynickaBirula:1970vy,Adler:1970gg,Papanyan:1973xa,Adler:1996cja,Schubert:2000yt,Gies:2016czm}.

For a given order $N$, the contribution $P_N$ is expressed in terms of $N$ Fourier integrals
\be\label{PNdef}
P_N=\int\ud\omega_1\dots\ud\omega_N\tilde{f}_1(\omega_1)\dots\tilde{f}_1(\omega_N)F_N(\omega_1,\dots,\omega_N) \;,
\ee
where $\tilde{f}_1$ is the Fourier transform of the weaker field and $F_N$ can be calculated using the worldline formalism, e.g. by starting with the master formula for the $N$-photon amplitude in~\cite{Schubert:2000yt}, see section~\ref{HigherOrderDerivation}. Again we approximate the strong field by a constant, which implies that $F_N$ contains a delta function $\delta(\omega_1+\dots+\omega_N)$ due to temporal homogeneity. This implies that some of the frequencies will be positive and the others negative. Without loss of generality we assume that $\omega_1$ to $\omega_J$ are the positive ones where $1\leq J<N$. It is convenient to introduce the normalized sum of these positive frequencies as
\be\label{SigmaSumomega}
\Sigma=\frac{1}{2m}\sum_{i=1}^J\omega_i \;.
\ee
Again using $m^2/(qE)$ as a large expansion parameter it can be shown, see section~\ref{HigherOrderDerivation}, that the dominant contribution to $F_N$ behaves as
\be\label{expWithoutFourier2}
F_N\sim\exp\left\{\!-\frac{2m^2}{qE}\left(\frac{\pi}{2}+i\phi(i\Sigma)\right)\right\}=\exp\left\{\!-\frac{2m^2}{qE}\Big(\arccos\Sigma-\Sigma\sqrt{1-\Sigma^2}\Big)\right\} \;.
\ee
Similar to the first order result~\eqref{firstorderW}, we recover the Schwinger exponent~\eqref{Sauter-Schwinger} for $\Sigma=0$, while the exponent vanishes for $\Sigma=1$, i.e. when the sum of the positive frequencies equals the mass gap of $2m$.

\subsection{Sauter pulse}

Now we are in the position to study different pulse shapes. For Sauter-like pulses, where the Fourier transform decays exponentially for large frequencies as in~\eqref{FourierSauterLargew}, we find that all orders $P_N$ have the same exponential behavior as our first order result~\eqref{P2-Sauter-integrated}
\be\label{PNSauter}
P_N^\text{Sauter}\sim\exp\left\{\!-\frac{2m^2}{qE}\left(\frac{\sqrt{\gamma_*^2-1}}{\gamma_*^2}+\arcsin\frac{1}{\gamma_*}\right)\right\} \;.
\ee
This also explains the good agreement between the first order result and the numerical simulation for small $\varepsilon$, since the higher-order terms come with extra powers of $\varepsilon$ but the same exponential. As expected, all the exponentials reduce to~\eqref{Sauter-Schwinger} at the threshold $\gamma_*\to1$. We can also verify the limit of large $\gamma_*$, where \eqref{PNSauter} reduces to the $E$-independent exponent in~\eqref{HighGammaSauter}, by expanding the exact result for a single Sauter pulse~\cite{NarozhnyiNikishovSimplest,Nikishov1985,Dunne:1998ni,Dunne:2004nc}.

\subsection{Gaussian pulse}

For a Gaussian pulse~\eqref{GaussFourier}, however, the situation is different. We calculate the $\omega_i$-integrals in~\eqref{PNdef} using the saddle point method, and find the dominant $\Sigma$ in analogy to~\eqref{domwGauss1},
\be\label{SGauss}
\Sigma_{\rm dom}=\left(1+\left[\frac{N}{2J(N-J)}\frac{qE}{\omega_{\rm fast}^2}\right]^2\right)^{-\frac{1}{2}} \;.
\ee
Inserting $\Sigma_{\rm dom}$ into the integrand of~\eqref{PNdef} we find that different orders have different exponentials. For even orders $N$ the dominant contribution comes from $J=N/2$, and for odd orders from $J=(N\pm1)/2$, and hence
\be\label{finNexpGauss}
\begin{split}
&N\text{ even:} \qquad P_N^{\rm Gauss}\sim\exp\left\{\!-\frac{2m^2}{qE}\arctan\left(\frac{2}{N}\frac{qE}{\omega_{\rm fast}^2}\right)\right\} \\
&N\text{ odd:} \qquad P_N^{\rm Gauss}\sim\exp\left\{\!-\frac{2m^2}{qE}\arctan\left(\frac{2N}{N^2-1}\frac{qE}{\omega_{\rm fast}^2}\right)\right\} \;.
\end{split}
\ee 
For $N=2$ this agrees with our result in~\eqref{firstorderGaussian}. As expected, all exponents in~\eqref{finNexpGauss} reduce to Schwinger's constant field result~\eqref{Sauter-Schwinger} as $\omega_{\rm fast}\to0$, and for $qE/\omega_{\rm fast}^2\ll1$ the strong field drops out leaving an exponential suppression given by $N$ factors of the Fourier transform~\eqref{GaussFourier} with $\omega=2m/(N/2)$ (only even orders remain without the strong field)
\be
P_N^{\rm Gauss}\left(\frac{qE}{\omega_{\rm fast}^2}\ll1\right)\sim\exp\left\{-N\left(\frac{2m}{N\omega_{\rm fast}}\right)^2\right\} \;.
\ee

The exponential suppression of $P_N$ is reduced at higher orders. However, due to the prefactor $\varepsilon^N$ there will in general be an order $N_{\rm dom}$ that yields the dominant contribution, in contrast to the Sauter case.  
To estimate this dominant order $N_{\rm dom}$, let us approximately treat $N$ as a continuum variable. Then we may estimate $N_{\rm dom}$ by the ``saddle point'' of $\varepsilon^N P_N$, which gives
\be\label{dominantNGauss}
N_{\rm dom}^{\rm Gauss}
\sim
\frac{2m^2}{qE}\,\frac{1}{|\ln\varepsilon|}\,
\frac{\sqrt{\chi^2-1}}{\chi^2}
\quad
{\rm where}
\quad
\chi
\sim
\frac{\gamma}{\sqrt{|\ln\varepsilon|}}
\sim
\frac{\gamma}{\gamma_{\rm crit}}
\;.
\ee
At threshold $\chi=1$ and for $\chi\gg1$, this dominant order vanishes indicating the
break-down of the continuum approximation for $N$. 
In these two cases, low orders of perturbation theory are sufficient.
In the region between the threshold $\gamma_{\rm crit}$ and $\gamma\gg\gamma_{\rm crit}$, however, 
the dominant order can be quite large $N_{\rm dom}^{\rm Gauss}\gg1$ if the 
electric field lies sufficiently below the critical value, 
$qE\ll m^2/|\ln\varepsilon|$. 
In this situation, low orders of perturbation theory are not enough and one 
has to use non-perturbative methods or estimate the higher orders in some 
other way.
This observation qualitatively explains the behavior in 
figure~\ref{GaussSpectra-fig}:
For the larger field (left plot), the zeroth plus first order amplitude is a good 
approximation while for the smaller field (right plot) strong deviations 
are visible.

Note that the above estimates are quite rough and only contain the 
leading order contributions (in $\varepsilon$ etc.). 
The accuracy of this estimate can be improved by taking into account further
contributions, such as factors of $E$ and $\omega_{\rm fast}$ accompanying $\varepsilon$ in $\varepsilon^N$.

Substituting the above value for the dominant order into~\eqref{finNexpGauss} we may estimate the total probability via
\be\label{PdominantOrder}
P_{e^+e^-}\sim\exp\left\{\!\frac{2m^2}{qE}i\phi\left(\frac{i\gamma_{\rm crit}}{\gamma}\right)\right\} \;.
\ee
Interestingly, this has formally the same form as in the Sauter case, provided we insert the corresponding critical value, which is $\gamma_{\rm crit}^{\rm Sauter}=\pi/2$ for a Sauter pulse but $\gamma_{\rm crit}^{\rm Gauss}\sim\sqrt{|\ln\varepsilon|}$ for a Gaussian pulse.

\subsection{Oscillating field}

As a third example we will consider a sinusoidal field, $f_1(t)=\cos(\omega_{\rm fast} t)$. For this field we have no nontrivial Fourier integrals to perform, so the exponent is directly given by~\eqref{expWithoutFourier2} with $\Sigma=N\omega_{\rm fast}/(4m)$. 
The threshold value for $\gamma$ where $\varepsilon^N P_N$ becomes larger than 
the zeroth order in~\eqref{Sauter-Schwinger} 
is given by
\be\label{threshold-sin}
\gamma_\text{cr}^{\rm cos}\sim|\ln\varepsilon| \;,
\ee
which agrees with the threshold found in~\cite{Linder:2015vta} by treating both fields nonperturbatively.
Estimating the dominant order with the ``saddle point'' for $N$, as above, we again find~\eqref{PdominantOrder} with $\gamma_{\rm cr}$ as in~\eqref{threshold-sin}. The dominant order is similar to the Gaussian case~\eqref{dominantNGauss},
\be
N_{\rm dom}^{\rm cos}
\sim
\frac{4m^2}{qE}\,\frac{1}{|\ln\varepsilon|}\,
\frac{\sqrt{\chi^2-1}}{\chi^2}
\quad
{\rm where}
\quad
\chi
\sim
\frac{\gamma}{|\ln\varepsilon|}
\sim
\frac{\gamma}{\gamma_{\rm crit}^{\rm cos}}
\;.
\ee
In the limit $\gamma\to\infty$ the strong field becomes negligible and we find
\be
\lim\limits_{\gamma\to\infty}N_{\rm dom}^{\rm cos}=\frac{4m}{\omega} \qquad P\sim\exp\left\{\!-\frac{4m^2}{qE \chi}\right\}\sim
\varepsilon^{4m/\omega} \;,
\ee
as expected (see e.g.~\cite{Dunne:2005sx}).

\section{Derivation of the first order spectrum}\label{Derivation first order spectrum}

In this section we will show how to derive the spectrum with a WKB-based formalism. For convenience and clarity, we will from now on absorb the charge into the background field $qE\to E$ and use units with $m=1$.  
We follow closely \cite{Hebenstreit:2011pm,Hebenstreit:2010vz} (see also~\cite{Kluger:1992gb}) in our treatment of the Dirac field. We use the ``Furry-picture'' where the spinor field solves the Dirac equation in the strong background field, $A^{\rm strong}_3(t)=A(t)$, 
\be
(i\slashed{\partial}-\gamma^3A(t)-1)\Psi(t,{\bf x})=0 \;.
\ee
The solution only depends nontrivially on time and is expressed in terms of two different sets of mode functions,
\be
\begin{split}
\Psi(t,{\bf x})&=\int\frac{\ud^3q}{(2\pi)^3}e^{-iq_ix^i}\sum\limits_{r=1,2}u_r(t,{\bf q})a_r({\bf q})+v_r(t,-{\bf q})b_r^\dagger(-{\bf q}) \\
&=\int\frac{\ud^3q}{(2\pi)^3}e^{-iq_ix^i}\sum\limits_{r=1,2}U_r(t,{\bf q})A_r(t,{\bf q})+V_r(t,-{\bf q})B_r^\dagger(t,-{\bf q}) \;,
\end{split}
\ee
where $a$ and $b^\dagger$ are the electron annihilation and positron creation operators, respectively, at $t\to-\infty$, and $A(\infty,{\bf q})$ and $B^\dagger(\infty,{\bf q})$ are the electron annihilation and positron creation operators, respectively, at $t\to+\infty$. The first set of mode functions are given by
\be\label{uandv}
\begin{split}
u_r(t,{\bf q})&=(i\gamma^0\partial_0+\gamma^i\pi_i(t,{\bf q})+1)g^+(t,{\bf q})R_r \\
v_r(t,-{\bf q})&=(i\gamma^0\partial_0+\gamma^i\pi_i(t,{\bf q})+1)g^-(t,{\bf q})R_r \;,
\end{split}
\ee
where $\pi_\LCperp=q_\LCperp$ ($=q_{1,2}$), $\pi_\spara=q_\spara-A_\spara(t)$ ($=\pi_3$), $\pi_0=\sqrt{1+\bm{\pi}^2}=:\sqrt{m_\LCperp^2+\pi_\spara^2}$, and
$g^\pm$ are the two solutions of the squared Dirac equation
\be
(\partial_0^2+\pi_0^2(t,{\bf q})+iA'(t))g^\pm(t,{\bf q})=0
\ee
that behave initially as
\be\label{Gpmdef}
\lim\limits_{t\to-\infty}g^\pm(t,{\bf q})\to[2\pi_0(\pi_0\pm\pi_\spara)]^{-\frac{1}{2}}\exp\bigg[\mp i\int\limits_{t_0}^t\!\ud t'\,\pi_0(t')\bigg]=:G^\pm(t,{\bf q}) \;.
\ee
This equation also gives the WKB approximations $G^\pm$, which we will use in the next section to calculate the first order amplitude. 
The spinors $R_r$, $r=1,2$ obey $\gamma^0\gamma^3 R_{1,2}=R_{1,2}$. We use the Weyl representation for the gamma matrices
\be
\gamma^0=\begin{pmatrix} 0&1_2 \\ 1_2&0 \end{pmatrix}
\qquad 
\gamma^i=\begin{pmatrix} 0&\sigma_i \\ -\sigma_i&0 \end{pmatrix} \;,
\ee   
where $\sigma_i$ are the Pauli matrices, 
and consequently
\be\label{R-spinors}
R_1=(0,\cos\varphi,\sin\varphi,0) \qquad
R_2=(0,-\sin\varphi,\cos\varphi,0)
\ee
for some $\varphi$. The second set of mode functions, the adiabatic ones, are chosen as \cite{Hebenstreit:2011pm,Hebenstreit:2010vz}
\be\label{UandV}
\begin{split}
U_r(t,{\bf q})&=(\gamma^0\pi_0+\gamma^i\pi_i+1)G^+(t,{\bf q})R_r \\
V_r(t,-{\bf q})&=(-\gamma^0\pi_0+\gamma^i\pi_i+1)G^-(t,{\bf q})R_r \;,
\end{split}
\ee
and the two sets of mode operators are related via a Bogoliubov transformation
\be\label{Bogoliubov}
\begin{split}
A_r(t,{\bf q})&=\alpha(t,{\bf q})a_r({\bf q})-\beta^*(t,{\bf q})b_r^\dagger(-{\bf q}) \\
B_r^\dagger(t,-{\bf q})&=\beta(t,{\bf q})a_r({\bf q})+\alpha^*(t,{\bf q})b_r^\dagger(-{\bf q}) \;,
\end{split}
\ee 
where $|\alpha(t,{\bf q})|^2+|\beta(t,{\bf q})|^2=1$.
The two sets satisfy the same commutation relations
\be
\{A_r(t,{\bf q}),A_{r'}^\dagger(t,{\bf q'})\}=\{a_r({\bf q}),a_{r'}^\dagger({\bf q'})\}=\delta_{rr'}(2\pi)^3\delta^3({\bf q}-{\bf q'})
\ee
and similarly for $b$ and $B$. 

We start in the in-vacuum state $\ket{0_\text{in}}$, and the final state contains an electron with momentum ${\bf p}$ and spin $s$ and a positron with momentum ${\bf p'}$ and spin $s'$. The amplitude for this process is to first order given by 
\be
\bra{0_\text{out}}B_{s'}(\infty,{\bf p'})A_s(\infty,{\bf p})\Big(1-i\int\!\ud t\;H_{\rm int}\Big)\ket{0_\text{in}}
=:(2\pi)^3\delta^3({\bf p}+{\bf p'})(\mathfrak{A}_0+\varepsilon\mathfrak{A}_1)\bracket{0_\text{out}}{0_\text{in}} \;,
\ee
where the interaction Hamiltonian given by~\eqref{interaction-Hamiltonian}. Using
\be
\bra{0_\text{out}}B_{s'}(\infty,{\bf p'})A_s(\infty,{\bf p})
=-\frac{\beta_p^*}{\alpha_p^*}\delta_{ss'}(2\pi)^3\delta^3({\bf p}+{\bf p'})\bra{0_\text{out}}+\frac{1}{\alpha_{-p'}^*\alpha_p^*}\bra{0_\text{out}}b_{s',p'}a_{s,p}
\;,
\ee
where $\beta_p=\beta(\infty,{\bf p})$, $\alpha_p=\alpha(\infty,{\bf p})\approx1$ and $\bracket{0_\text{out}}{0_\text{in}}\approx1$, we find
\be\label{M0frombeta}
\mathfrak{A}_0=-\delta_{ss'}\beta_p^*  \;,
\ee
\be\label{M1Def}
\varepsilon\mathfrak{A}_1=-i\int\!\ud t\;\bar{u}_{s,{\bf p}}(t)\slashed{A}^{\rm fast}(t)v_{s',-{\bf p}}(t) \;.
\ee

\subsection{Zeroth order}\label{Zeroth order}

To obtain the zeroth order amplitude~\eqref{M0frombeta} we need $\beta$, which can be obtained from the $g^\LCp$ solution, see e.g.~\cite{Nikishov1985,Hebenstreit:2011pm,Hebenstreit:2010vz}.
In a constant electric field $A=Et$ one finds, see e.g.~\cite{Hebenstreit:2011pm,Hebenstreit:2010vz,NikishovConstant}, 
\be\label{gp-constant}
g^\LCp_{\rm const}=\frac{1}{\sqrt{2E}}\exp\Big\{\frac{i\eta}{4}\Big(1+\ln\frac{2}{\eta}\Big)-\frac{i\pi}{4}-\frac{\pi\eta}{8}-\frac{i\eta}{2}\phi\Big(\frac{p_\spara-Et_0}{m_\LCperp}\Big)\Big\}D_{\!-1+\frac{i\eta}{2}}\big(\!-e^{-\frac{i\pi}{4}}u\big) \;,
\ee
where $\eta=\frac{m_\LCperp^2}{E}$ and $u=-\sqrt{\frac{2}{E}}(p_\spara-Et)$.
The phase $\phi$ is given by~\eqref{phi-function}.
Usually a constant phase is irrelevant, but here we need to be a little more careful to obtain the correct relative phase between the zeroth and first order terms, which is important for the interference in the momentum spectrum between $\mathfrak{A}_0$ and $\mathfrak{A}_1$. It follows from the $t\to\infty$ limit of~\eqref{gp-constant} that
\be
\alpha_{\rm const}=\frac{\sqrt{\pi\eta}}{\Gamma\big(1-\frac{i\eta}{2}\big)}\exp\Big\{\frac{i\eta}{2}\Big(1+\ln\frac{2}{\eta}\Big)-\frac{i\pi}{4}-\frac{\pi\eta}{4}\Big\}\approx1
\ee
and
\be\label{betaconstantE}
\beta_{\rm const}=-\exp\left(-\frac{\pi m_\LCperp^2}{2E}-\frac{i  m_\LCperp^2}{E}\phi\left[\frac{p_\spara-Et_0}{m_\LCperp}\right]\right) \;.
\ee  

For a Sauter pulse, $A=\frac{E}{\sigma}\tanh\sigma t$, one finds, see e.g.~\cite{Hebenstreit:2011pm,Hebenstreit:2010vz},
\be
g^\LCp_{\rm Sauter}=e^{i\theta_1}\frac{u^{-\frac{i\pi_{\scriptscriptstyle0}^\text{\tiny in}}{2\sigma}}(1-u)^\frac{i\pi_{\scriptscriptstyle0}^\text{\tiny out}}{2\sigma}}{\sqrt{2\pi_{\scriptscriptstyle0}^\text{\tiny in}(\pi_{\scriptscriptstyle0}^\text{\tiny in}+\pi_\spara^\text{\tiny in})}}{}_2F_1(a,b,c,u) \qquad u=\frac{1}{2}(1+\tanh\sigma t) \;,
\ee
where ``in'' and ``out'' refer to $t\to-\infty$ and $t\to+\infty$, respectively, and the parameters in the hypergeometric function are
\be
a=1+\frac{iE}{\sigma^2} \qquad b=-\frac{iE}{\sigma^2}-i\frac{\pi_{\scriptscriptstyle0}^\text{\tiny in}-\pi_{\scriptscriptstyle0}^\text{\tiny out}}{2\sigma} \qquad c=1-i\frac{\pi_{\scriptscriptstyle0}^\text{\tiny in}}{\sigma} \;.
\ee 
The asymptotic Bogoliubov coefficients obtained from $g^\LCp$ are given by
\be\label{alphabetaSauter}
\begin{split}
\alpha_{\rm Sauter}&=e^{i(\theta_1+\theta_2)}\sqrt{\frac{\pi_{\scriptscriptstyle0}^\text{\tiny out}(\pi_{\scriptscriptstyle0}^\text{\tiny out}+\pi_\spara^\text{\tiny out})}{\pi_{\scriptscriptstyle0}^\text{\tiny in}(\pi_{\scriptscriptstyle0}^\text{\tiny in}+\pi_\spara^\text{\tiny in})}}\frac{\Gamma(c)\Gamma(c-a-b)}{\Gamma(c-a)\Gamma(c-b)} \\
\beta_{\rm Sauter}&=e^{i(\theta_1-\theta_2)}\sqrt{\frac{\pi_{\scriptscriptstyle0}^\text{\tiny out}(\pi_{\scriptscriptstyle0}^\text{\tiny out}-\pi_\spara^\text{\tiny out})}{\pi_{\scriptscriptstyle0}^\text{\tiny in}(\pi_{\scriptscriptstyle0}^\text{\tiny in}+\pi_\spara^\text{\tiny in})}}\frac{\Gamma(c)\Gamma(a+b-c)}{\Gamma(a)\Gamma(b)} \;,
\end{split}
\ee 
where the phases are given by
\be
\theta_1=\pi_{\scriptscriptstyle0}^\text{\tiny in}t_{\scriptscriptstyle0}+\int\limits_{-\infty}^{t_{\scriptscriptstyle0}}\!\ud t'(\pi_{\scriptscriptstyle 0}-\pi_{\scriptscriptstyle0}^\text{\tiny in}) \qquad \theta_2=-\pi_{\scriptscriptstyle0}^\text{\tiny in}t_{\scriptscriptstyle0}+\int\limits_{t_{\scriptscriptstyle0}}^\infty\!\ud t'(\pi_{\scriptscriptstyle 0}-\pi_{\scriptscriptstyle0}^\text{\tiny out}) \;.
\ee
Again these phases are important for the interference between $\mathfrak{A}_0$ and $\mathfrak{A}_1$.

\subsection{First order}\label{FirstOrderDerivation}

To calculate the first order amplitude~\eqref{M1Def} we begin by approximating the exact wave functions with the WKB approximations $U$ and $V$ in~\eqref{UandV}. We Fourier transform the weak field according to
\be\label{Fouriertransformationdef}
f(t)=\int\frac{\ud\omega}{2\pi} e^{-i\omega t}\tilde{f}(\omega) \;.
\ee
We perform the time integral with the saddle point method,
\be\label{firstorder-timeintegral}
\int\!\ud t\;\frac{m_\LCperp}{\pi_{\scriptscriptstyle0}}\exp\Big\{\!-i\omega t+2i\int_0^t\!\ud t'\pi_{\scriptscriptstyle0}(t')\Big\}=\frac{m_\LCperp}{\pi_{\scriptscriptstyle0}}\Big(\frac{i\pi}{\dot{\pi}_{\scriptscriptstyle0}}\Big)^\frac{1}{2}\exp\Big\{\!-i\omega t+2i\int_0^t\!\ud t'\pi_{\scriptscriptstyle0}(t')\Big\} \;,
\ee
where the saddle point $t(\omega)$ is obtained from 
\be\label{timeSaddleEq}
2\pi_{\scriptscriptstyle0}(t)=\omega \;.
\ee
If we assume that the strong field is essentially constant in the region where the weak field is effectively nonzero, then the integral in the exponent has a simple form
\be
2i\int_0^t\!\ud t'\pi_{\scriptscriptstyle0}(t')=-\frac{im_\LCperp^2}{E}\Big(\phi\Big[\frac{\pi_\spara}{m_\LCperp}\Big]-\phi\Big[\frac{p_\spara}{m_\LCperp}\Big]\Big) \;.
\ee
This equation together with~\eqref{firstorder-timeintegral} and the saddle point from~\eqref{timeSaddleEq} lead to~\eqref{exponentinW1}.

\subsubsection{Sauter-like pulses}

We begin with a weak field in the shape of a Sauter pulse with Fourier transform~\eqref{FourierSauterExact}.
We assume that, although the weak field is rapidly varying, we still have $\omega_\text{fast}\ll1$, so we can approximate the Fourier transform as in~\eqref{FourierSauterLargew}. The exponential decay of~\eqref{FourierSauterLargew} is the most important property of the weak field, and other fields with similar Fourier transforms lead to similar calculations. With such similar fields in mind, we write for convenience $\omega_*:=E\gamma_*$, so the Fourier transform decays as $e^{-|\omega|/\omega_*}$ ($\omega_*=2\omega/\pi$ for a Sauter pulse). 
Performing the Fourier integral with the saddle point method gives (using~\eqref{timeSaddleEq})
\be\label{firstorder-SauterFourierInt}
\int\!\ud\omega\;...\exp\Big\{\!-\frac{\omega}{\omega_*}-i\omega t+2i\int_0^t\!\ud t'\pi_{\scriptscriptstyle0}(t')\Big\}=...\Big(\frac{4\pi\dot{\pi}_{\scriptscriptstyle0}}{i}\Big)^\frac{1}{2}\exp\Big\{2i\int_0^t\!\ud t'\pi_{\scriptscriptstyle0}(t')\Big\} \;,
\ee
where the ellipses stand for terms in the prefactor that vary slowly compared to the exponential, and where the saddle point equation is $t(\omega_{\rm dom})=i/\omega_*$, which, together with~\eqref{timeSaddleEq}, gives
\be\label{tandwSaddlesSauter}
t=\frac{i}{\omega_*} \qquad \omega_{\rm dom}=2\pi_{\scriptscriptstyle0}\Big(t=\frac{i}{\omega_*}\Big) \;.
\ee
The $\dot{\pi}$-terms in~\eqref{firstorder-timeintegral} and~\eqref{firstorder-SauterFourierInt} cancel and we find  
\be\label{M1SauterSauter}
\varepsilon\mathfrak{A}_1=\delta_{ss'}2\pi\frac{E\varepsilon}{\omega_\text{fast}^2}\frac{m_\LCperp}{\pi_{\scriptscriptstyle0}}
\exp\Big\{2i\int_0^t\!\ud t'\pi_{\scriptscriptstyle0}(t')\Big\} \;,
\ee
which for a constant strong field becomes (c.f.~\eqref{frakA1result})
\be\label{M1result}
\varepsilon\mathfrak{A}_1=\delta_{ss'}2\pi\frac{E\varepsilon}{\omega_{\rm fast}^2}\frac{m_\LCperp}{\pi_{\scriptscriptstyle0}}
\exp\Big\{\!-\frac{im_\LCperp^2}{E}\Big(\phi\Big[\frac{\pi_\spara}{m_\LCperp}\Big]-\phi\Big[\frac{p_\spara}{m_\LCperp}\Big]\Big)\Big\} \;,
\ee 
where
\be
\pi_\spara=p_\spara-\frac{i}{\gamma_*} \qquad \pi_{\scriptscriptstyle0}=\sqrt{m_\LCperp^2+\Big(p_\spara-\frac{i}{\gamma_*}\Big)^2} \;.
\ee
Note that \eqref{betaconstantE} and \eqref{M1result} contain a phase that comes from $t_0=0$, which cancels in the probability. 

\subsubsection{Gaussian spectrum}

For a weak Gaussian field we find after performing the time and Fourier integrals using the saddle point method
\be\label{M1DefGauss}
\varepsilon\frak{A}_1=\delta_{ss'}\frac{E\varepsilon\sqrt{\pi}}{2m_\LCperp\omega_{\rm fast}}\frac{1}{\Sigma^2}\left[1+\nu^2+\frac{iP\nu}{\Sigma}\right]^{-\frac{1}{2}}\exp\left\{-\frac{m_\LCperp^2}{E}\left[iP\Sigma+\arccos\Sigma-i\phi(P)\right]\right\} \;,
\ee
where $\Sigma$ is the normalized Fourier frequency at the saddle point
\be
\Sigma=\frac{\omega}{2m_\LCperp}=\frac{\sqrt{1+\nu^2+P^2}-iP\nu}{1+\nu^2} \;,
\ee
where $P=p_3/m_\LCperp$ and $\nu=E/\omega_{\rm fast}^2$. This saddle point reduces to the dominant frequency in~\eqref{domwGauss1} for $\f{p}=0$.
Squaring~\eqref{M1DefGauss} and performing the momentum integrals with the saddle point method gives~\eqref{firstorderGaussian}.

\section{Derivation of higher orders}\label{HigherOrderDerivation}

Expanding the effective action in the field strength of the weak field and Fourier transforming it, we obtain an expression, see~\eqref{PNdef}, that corresponds to the amplitude of $N$ off-shell photons interacting in the strong constant electric field via a dressed fermion loop. There are general and compact ``master formulas'' for N-photon amplitudes in a constant electromagnetic field, derived using the worldline formalism~\cite{Shaisultanov:1995tm,Schubert:2000yt}. These formulas offer one way to obtain the higher orders $P_N$ in~\eqref{Pseriesdef}.
Starting with eq.~3.13 in~\cite{Schubert:2000yt}, which holds for arbitrary constant electromagnetic fields and off-shell photons, in particular for a constant electric field and for photons with $k_{i,\mu}=(\omega_i,0,0,0)$, $i=1,...,N$. We begin by replacing the polarization vectors $\epsilon_{i,\mu}$ in~\cite{Schubert:2000yt} with the Fourier transform of the weak field\footnote{The probability is given by the imaginary part of the amplitude.}. This leads to an expression for the $N$-th order $\varepsilon^N P_N$ as in~\eqref{PNdef}, where $F_N$ is given by 
\be\label{MasterFormulaStart}
F_N\sim\delta\Big(\sum\limits_{i=1}^N\omega_i\Big)\int_0^\infty\ud T\int_0^1\prod\limits_{i=1}^N\ud\tau_i
\dots\exp\Big\{-\frac{1}{E}\Big(T+\frac{1}{2}\sum\limits_{i,j=1}^N\omega_iE\mathcal{G}_{i,j}\omega_j\Big)\Big\} \;,
\ee 
where $\mathcal{G}_{i,j}:=\bar{\mathcal{G}}_B^{4,4}(\tau_i,\tau_j)$ is the ``time-time'' component of the worldline propagator, which can be obtained from eq.~3.14 and 3.7 in~\cite{Schubert:2000yt},  
\be
E\mathcal{G}_{i,j}=\frac{\cos(T\dot{G}_{i,j})-\cos T}{2\sin T}
\qquad
\dot{G}_{i,j}=\text{sign}(\tau_i-\tau_j)-2(\tau_i-\tau_j) \;.
\ee
The ellipses in~\eqref{MasterFormulaStart} stand for pre-exponential terms that depend on the integration variables but do not affect the exponential part of the probability. Here we content ourselves with the exponential, which allows us to understand why higher orders can be important.   
We obtain the dominant contribution at each order $N$ by looking for the values of the integration variables that maximize the exponential. This is similar to the saddle point approximation, except that not all of these integrals are Gaussian around the maximum. We begin with the $\tau_i$-integrals. We note that $-1\leq\epsilon(\tau_i-\tau_j)\dot{G}_{i,j}\leq1$ and assume that $0<T<\pi$ (this is consistent with the saddle points for $T$ obtained below), which imply 
$0\leq E\mathcal{G}_{i,j}\leq\frac{1}{2}\tan\frac{T}{2}=:E\mathcal{G}$.
For terms in~\eqref{MasterFormulaStart} with $\omega_i\omega_j>0$ the exponential is maximized by $E\mathcal{G}_{i,j}=0$, which is obtained with $|\tau_i-\tau_j|=0,1$, and for $\omega_i\omega_j<0$ the exponential is maximized by $E\mathcal{G}_{i,j}=E\mathcal{G}$, which is obtained with $|\tau_i-\tau_j|=1/2$. 
We number the $\tau$'s such that $E\mathcal{G}_{i,j}=0$ for $i,j=1,...,J<N$ or $i,j=J+1,...,N$, and $E\mathcal{G}_{i,j}=E\mathcal{G}$ for $i=1,...,J$ and $j=J+1,..,N$, where $J$ characterizes the saddle point and will be determined below. The values of $\omega_i$ that give the dominant contribution, which we find below, agree with the assumption $\omega_i\omega_j<0$ for $i=1,...,J$ and $j=J+1,..,N$. Using the delta function for $\omega_i$, we obtain for the dominant values of $\tau_i$, 
\be\label{expWithoutFourier}
\exp\Big\{\!-\frac{1}{E}\Big(T+E\mathcal{G}\sum\limits_{i=1}^J\omega_i\!\sum\limits_{j=J+1}^N\!\omega_j\Big)\Big\} =\exp\Big\{\!-\frac{2}{E}\Big(\frac{T}{2}-\Sigma^2\tan\frac{T}{2}\Big)\Big\} \;,
\ee
where $\Sigma$ is given by (c.f.~\eqref{SigmaSumomega})
\be\label{Sigmadefagain}
\Sigma=\frac{1}{2}\sum_{i=1}^J\omega_i \;.
\ee
We assume without loss of generality that $\Sigma>0$.
With $0<\Sigma<1$, the saddle point for $T$ is given by $T=2\arccos\Sigma$, which agrees with our assumption $0<T<\pi$. With this saddle point we obtain the exponential contribution (c.f.~\eqref{expWithoutFourier2})
\be\label{expWithoutFourier2again}
F_N\sim\exp\left\{\!-\frac{2}{E}\Big(\arccos\Sigma-\Sigma\sqrt{1-\Sigma^2}\Big)\right\} \;.
\ee

\subsection{Sauter-like pulses}

By Sauter-like fields we mean fields with Fourier transforms decaying as
\be\label{exponential-Fourier}
\tilde{f}_1(\omega)\sim e^{-|\omega|/\omega_*} \;.
\ee
For example, for a Sauter pulse we have $\omega_*=2\omega_{\rm fast}/\pi$. For these types of fields we have
\be\label{exp0Exponential}
\tilde{f}_1(\omega_1)\dots\tilde{f}_1(\omega_N)\sim\exp\Big\{\!-\frac{1}{E}\sum\limits_{i=1}^N\frac{|\omega_i|}{\gamma_*}\Big\} \;,
\ee
where $\gamma_*=\omega_*/E$. The exponential is maximized for $\Sigma>0$ by
$\omega_i\geq0$ and $\omega_j\leq0$ (and for $\Sigma<0$ by $\omega_i\leq0$ and $\omega_j\geq0$) with $i=1,...,J$, $j=J+1,...,N$, where
\be
\tilde{f}_1(\omega_1)\dots\tilde{f}_1(\omega_N)\sim\exp\Big\{\!-\frac{1}{E}\frac{4|\Sigma|}{\gamma_*}\Big\} \;.
\ee
Note that this does not depend on $J$ or $N$, which, since~\eqref{expWithoutFourier2again} also only depends on $\Sigma$, implies that all orders have the same exponential for these Sauter-like fields. Assuming without loss of generality that $\Sigma>0$, the saddle point is given by
$\Sigma=\sqrt{\gamma_*^2-1}/\gamma_*$, 
which satisfies $0<\Sigma<1$ for $\gamma_*>1$.
With this saddle point for $\Sigma$ we finally find~\eqref{PNSauter}, i.e. the same exponential that we found to first order, $P_2$, and which one also finds by treating the weak field nonperturbatively, e.g. with worldline instantons.

\subsection{Gaussian pulse}

The Fourier transform of a Gaussian pulse is given by~\eqref{GaussFourier}.
We use the delta function to perform the $\omega_N$-integral and change variable from $\omega_1$ to $\Sigma$ as defined in~\eqref{Sigmadefagain}, which yields
\be
\tilde{f}_1(\omega_1)\dots\tilde{f}_1(\omega_N)\sim\exp\Big\{\!-\frac{1}{E}\frac{1}{4\lambda^2}\Big(\Big[2\Sigma-\sum\limits_{i=2}^J\omega_i\Big]^2+\sum\limits_{i=2}^{N-1}\omega_i^2+\Big[2\Sigma+\!\sum\limits_{i=J+1}^{N-1}\!\omega_i\Big]^2\Big)\Big\}
\;,
\ee
where $\lambda=\omega_{\rm fast}/\sqrt{E}$. The saddle points for the $\omega$-integrals are $\omega_i=2\Sigma/J$ for $i=2,...,J$ and $\omega_j=-2\Sigma/(N-J)$ for $j=J+1,...,N-1$ (notice $\omega_i\omega_j<0$), which give
\be
\tilde{f}_1(\omega_1)\dots\tilde{f}_1(\omega_N)\sim\exp\Big\{\!-\frac{2}{E}\nu\Sigma^2\Big\} 
\qquad 
\nu=\frac{N}{2J(N-J)}\frac{1}{\lambda^2} \;.
\ee
The saddle points for $\Sigma$ and $T$ are given by (c.f.~\eqref{SGauss})
$\Sigma=\frac{1}{\sqrt{1+\nu^2}}$
and
$T=2\arctan\nu$,
which satisfy $0<\Sigma<1$ and $0<T<\pi$. This gives 
\be\label{finNJexpGauss}
P_N\sim\exp\Big\{\!-\frac{2}{E}\arctan\nu\Big\} \;.
\ee 
The dominant saddle point is given by $J=N/2$ for even $N$, and $J=(N\pm1)/2$ for odd $N$, and hence we finally find~\eqref{finNexpGauss}.

\section{Numerical calculation of the pair production probability}
\label{numerics}

To check the validity of the approximations used, we compare the
analytical results to a numerical evaluation of the pair production
probability. 
One option would be to use the quantum kinetic formalism~\cite{Schmidt:1998vi} (see~\cite{Hebenstreit:2011pm} and references therein) as in~\cite{Orthaber:2011cm,Otto:2014ssa,Panferov:2015yda}. We have instead used an equivalent~\cite{Dumlu:2009rr,Dunne:2008kc} method based on the Riccati equation~\cite{Dumlu:2011rr,Schneider:2016vrl}.
The two different sets of mode functions~\eqref{uandv}
and~\eqref{UandV}, together with the transformation between the
particle creation and annihilation operators~\eqref{Bogoliubov}, give
a relation between the Bogoliubov coefficients $\alpha(t, {\bf q})$
and $\beta(t, {\bf q})$ that can be turned into a time evolution
equation for the ratio
$R(t, {\bf q}) = \beta(t, {\bf q})/\alpha(t, {\bf q})$:\footnote{Note
  that in this case we assume that the spinor field is a solution of
  the Dirac equation in the full field
  $A_3(t) = A_3^\text{strong} + A_3^\text{fast}$ instead
  of the strong field alone.}
\begin{align}
  \label{Riccati}
  \begin{split}
    \dot{R}(t, {\bf q}) &= \Xi(t, {\bf q}) \left(
      e^{2 i \phi(t, {\bf q})}
      + R^2(t, {\bf q}) e^{- 2 i \phi(t, {\bf q})}
    \right) \;, \\
    \Xi(t, {\bf q}) &= \frac{\dot{A}(t) m_\perp}{2 \pi_0^2(t,{\bf q})} \;,\\
    \phi(t, {\bf q}) & = \int_{t_0}^t\!\ud t'\, \pi_0(t',{\bf q}) \;,
  \end{split}
\end{align}
which is a Riccati equation; for a detailed derivation
see~\cite{Dumlu:2011rr}. To numerically integrate~\eqref{Riccati} we
proceed as in~\cite{Schneider:2016vrl}, replacing the integral expression
for $\phi(t, {\bf q})$ by the differential equation
$\dot{\phi}(t, {\bf q}) = \pi_0(t,{\bf q})$ and solve for $R$ and $\phi$
in lockstep. To combat numerical instabilities due to the oscillatory
nature of~\eqref{Riccati} and the very small quantities involved, we
employ the \textsc{tides} package~\cite{Tides} that uses the
\textsc{gnu mpfr} library~\cite{Mpfr} for multiple precision
arithmetic to integrate the Riccati equation over a time interval
$[-T, T]$ (for more information see~\cite{Schneider:2016vrl}). In
practice, to choose the integration region $[-T, T]$ and number of
digits to use in the calculation, both were increased until the result
does not change significantly, using the analytic solution for a
single Sauter pulse as a benchmark.

\section{Conclusions}\label{Conclusions}

In order to understand the dependence of the dynamically assisted 
Sauter-Schwinger effect on the shape of the weaker field $f_1(t)$, we employ a 
perturbative expansion in terms of the weaker field while treating the strong 
field $f_0(t)$ non-perturbatively.
It turns out that this dependence can be understood in terms of the Fourier 
transform $\tilde{f}_1(\omega)$, especially its decay for large frequencies.
In case of exponential decay (such as for a Sauter pulse), the higher orders
of this perturbative expansion display the same exponential behavior as the 
first order amplitude -- such that the zeroth plus first order amplitudes are 
sufficient in general.
For faster (e.g. Gaussian) decay at large frequencies, however, the various
orders display different exponential behaviors and -- depending on the 
parameters -- higher orders can yield the dominant contribution. Further, already the first order contribution explains the different behavior of the threshold, $\gamma_{\rm crit}^{\rm Sauter}\approx\pi/2$ versus $\gamma_{\rm crit}^{\rm Gauss}\sim\sqrt{|\ln\varepsilon|}$.
We compared our findings to numerical simulations as well as previous results
and showed that they can nicely be understood in this picture.
As an outlook, it would be interesting to consider non-analytic functions 
$f_1(t)$ with a slower-than-exponential decay~\cite{harmonicAnalysisBook}
such as fields with compact support, similar to those considered in 
e.g.~\cite{Otto:2014ssa,Mocken:2010uhp,Aleksandrov:2017owa}.

Note that the approach used here displays some similarities to other derivations
of pair creation by combinations of two fields, where one is treated 
non-perturbatively and the other one perturbatively. For example, the stronger field could be a plane wave, which can be treated exactly using Volkov 
solutions, while the weaker field could be another plane wave or even a single 
photon (Breit-Wheeler type processes, see e.g.~\cite{Heinzl:2010vg,Jansen:2013dea,Meuren:2015mra,Nousch:2015pja} for some recent studies), the Coulomb field of a nucleus 
(Bethe-Heitler type processes, see e.g.~\cite{Yakovlev-Volkov-perturbativeCoulomb,Milstein:2006zz,DiPiazza:2009py,
Augustin:2014xga}), or other shapes, such as a delta-function pulse~\cite{Fedotov:2013uja}.  
However, depending on the character of the two fields, the physical behavior
can be quite different.
For example, a plane wave alone cannot 
produce pairs, such that there is no zeroth-order contribution and thus also
no interference between zeroth and first order amplitudes. 
The interaction of a single (on-shell) photon with a strong and slowly varying or even 
constant electric field (see e.g.~\cite{Dunne:2009gi}) is also different 
from the scenario considered here, where the weaker field is purely 
time-dependent, because the constraints from energy-momentum conservation are 
quite different. 
Recently, we studied the combined impact of these three fields, i.e., 
a strong and slowly varying field plus a weaker time-dependent field plus a 
high-energy photon 
(``doubly assisted'' Sauter-Schwinger effect~\cite{Torgrimsson:2016ant}) 
and found further enhancement (see also~\cite{Jansen:2013dea,Nousch:2015pja} for photon-stimulated pair production in bi-frequent plane waves). 
In this work~\cite{Torgrimsson:2016ant}, the strong and weaker field were both
treated non-perturbatively. 
Whether and when the weaker field can also be treated via the perturbative 
approach presented here will be the subject of further studies~\cite{DoublyAssisted2}.

The remarkable agreement in Figures~\ref{SauterSpec033-fig} 
and~\ref{TanhSech033-fig} suggests that the approach presented here is a 
quite powerful method for studying a large class of time-dependent fields
by means of closed analytical expressions.  
For example, considering a strong and slowly varying field plus a 
superposition of several weaker Sauter pulses, the first-order amplitude is 
just given by the sum of the amplitudes for single Sauter pulses 
(with the associated phases etc.), which could then be used to study the 
momentum spectra in the combination of multiple-slit interference 
effects~\cite{Akkermans:2011yn} and dynamical assistance, 
c.f.~\cite{Li:2014psw}.

In this paper we have focused on fields depending only on time and having only one nonzero component. This is partly because of the numerical method we have chosen for checking our analytical approximations with the exact result. Our analytical approach can for example be generalized to weak fields that are not parallel to the strong field. In appendix~\eqref{Higher orders from worldline instantons} we consider a time-dependent weak field but a spatially inhomogeneous strong field. Pair production by a strong, spatially inhomogeneous Sauter pulse and a weak, time-dependent Sauter pulse was studied in~\cite{Schneider:2014mla}, where the exponential part of the pair production probability was obtained by treating both fields nonperturbatively. We show in appendix~\eqref{Higher orders from worldline instantons} that this exponential can also be obtained by treating the weak field perturbatively. Given that we have shown that, in the spatially homogeneous case, the perturbative approach also leads to a good approximation for the prefactor, appendix~\eqref{Higher orders from worldline instantons} seems like a promising starting point for further studies of pair production by fields depending on both time and space, where the probability can be very sensitive to the field shape in dynamical assistance~\cite{Linder:2015vta} while spatially inhomogeneous fields exhibit universal features in the probability~\cite{Gies:2015hia,Gies:2016coz}.

\appendix

\section{First order from the polarization tensor}\label{First order from the polarization tensor}

We can also obtain the total probability $P_2$ using the exact polarization tensor in a constant electric field~\cite{Dittrich:2000zu}. This is similar to the use in e.g.~\cite{Milstein:2006zz,DiPiazza:2009py} of the exact polarization tensor in a plane wave to obtain the probability of pair production by a combination of a strong laser field and a Coulomb field that is treated to lowest order. The exact expression for the polarization tensor $\Pi^{\mu\nu}$ in a constant electric field can be found in eq.~2.59 in~\cite{Dittrich:2000zu}. With a weak field in the form of~\eqref{Fouriertransformationdef} we need $\Pi^{33}$ evaluated with $k^\mu=(\omega,{\bf 0})$. One finds\footnote{The contact term does not contribute here.}
\be
\Pi^{33}=-\frac{\alpha}{2\pi}\omega^2\int\limits_0^\infty\frac{\ud s}{s}\int\limits_{-1}^1\frac{\ud v}{2}\frac{s}{\sinh s}Ne^{-\frac{is\phi_0}{E}} \;,
\ee  
where
\be
N=N_0+N_1=2\frac{\cosh s-\cosh vs}{\sinh^2s}
\qquad
\phi_0=1-\frac{\omega^2}{2}\frac{\cosh s-\cosh vs}{s\sinh s} \;.
\ee
The second order contribution to the pair production probability $\varepsilon^2P_2$ is obtained by expanding the effective action to second order in the weak field
\be
\varepsilon^2P_2=2\text{Im }\Gamma[A(t)=Et+A_{\rm fast}(t)]\Big|_{A_{\rm fast}^2}\sim\text{Im }A_{\rm fast}\Pi[Et] A_{\rm fast} \;,
\ee
where the factor of $2$ disappears because the polarization tensor is related to the effective action via 
\be
\epsilon\Pi\epsilon'=\Gamma[A_\mu+\epsilon_\mu e^{-ikx}+\epsilon'_\mu e^{-ik'x}]\Big|_\text{linear in $\epsilon$ and $\epsilon'$} \;.
\ee 
Since the strong field is constant the Fourier frequency of the weak field is conserved, and so
\be\label{P2fromPolarizationGeneral}
\varepsilon^2P_2=\text{Im}\int\frac{\ud\omega_1}{2\pi}\frac{\ud\omega_2}{2\pi}a(\omega_1)a(\omega_2)2\pi\delta(\omega_1+\omega_2)\Pi^{33}
=\text{Im}\int\frac{\ud\omega}{2\pi}|a(\omega)|^2\Pi^{33} \;.
\ee
With the Fourier transform for the Sauter pulse~\eqref{FourierSauterLargew} and Gaussian pulse~\eqref{GaussFourier}, we perform all the integrals in~\eqref{P2fromPolarizationGeneral} with the saddle point method, and after some straightforward calculations we recover~\eqref{P2-Sauter-integrated} and~\eqref{firstorderGaussian}.

\section{Higher orders from worldline instantons}\label{Higher orders from worldline instantons}

In this section we will 1) give an alternative derivation of our higher order results~\eqref{expWithoutFourier2}, and 2) generalize to a spatially inhomogeneous strong field. We will, in particular, rederive the results in~\cite{Schneider:2014mla} for dynamical assistance in a double Sauter pulse obtained by replacing $t$ with $z$ in the strong field in~\eqref{double-Sauter}. For this we will use worldline instantons, but, as we continue to treat the weak field perturbatively using its Fourier transform, these worldline instantons are different from those where both the strong and the weak field are treated nonperturbatively as in~\cite{Schneider:2014mla}. We start with the worldline representation of the effective action (for scalar QED for simplicity), see e.g.~\cite{Dunne:2005sx},
\be
\text{Im }\Gamma_N=\text{Im}\int\limits_0^\infty\frac{\ud T}{T}\oint\mathcal{D}x\Big(\!-i\int_0^1\!a(t)\dot{z}\Big)^N
\exp-i\Big(\frac{T}{2}+\int_0^1\frac{\dot{x}^2}{2T}+A\dot{x}\Big) \;.
\ee
We separate the zero modes from the instanton, $x(\tau)\to x_c+x(\tau)$, where 
\be
\int_0^1 x(\tau)=0 \;.
\ee
The integral over $t_c$ gives
$\delta(\omega_1+...+\omega_N)$.
Given the derivation in section~\ref{HigherOrderDerivation}, we assume $\tau_i=\tau_1$ for $i=1,...,J$ and $\tau_i=\tau_N$ for $i=J+1,...,N$. With $\Sigma$ as in~\eqref{SigmaSumomega}, the instanton exponent is given by
\be\label{instanton-exp-1}
\exp-i\Big(2\Sigma[t(\tau_1)-t(\tau_N)]+T+\int_0^1A\dot{x}\Big) \;,
\ee  
where the $T$-integral has been performed with the saddle point method,
$T^2=\dot{t}^2-\dot{z}^2$
(there is no step function in $T$ because of symmetry). We want to compare with results in~\cite{Schneider:2014mla}, so we choose a spatially inhomogeneous strong field, $A_0(z)$. 
The instanton equations are given by
\be
\ddot{z}=-TA_0'\dot{t} \qquad
\ddot{t}=-TA_0'\dot{z}+2\Sigma T[\delta(\tau-\tau_1)-\delta(\tau-\tau_N)] \;.
\ee
The second equation immediately gives us
\be
\dot{t}=-T\Big(A_0-2\Sigma[\theta(\tau-\tau_1)+\tau_1-\theta(\tau-\tau_N)-\tau_N]\Big) \;,
\ee
where we have assumed that the field is antisymmetric so that the $\tau$-integral of $A_0$ vanishes. 
With $\tau_N=0$ and $\tau_1=1/2$ we obtain
\be
\dot{t}=-T\left[A_0-\Sigma\text{ sign}\left(\tau-\frac{1}{2}\right)\right]
\qquad
\dot{z}=\pm iT\sqrt{1-\left[A_0-\Sigma\text{ sign}\left(\tau-\frac{1}{2}\right)\right]^2} \;.
\ee
Using partial integration and the instanton equations the instanton exponent~\eqref{instanton-exp-1} becomes
\be\label{general-z-int}
F_N=\exp\Big\{\frac{i}{T}\int_0^1\dot{z}^2\Big\}
=\exp\Big\{\!-4\int_0^{\hat{z}}\ud z\sqrt{1-\Big(A_0(z)+\Sigma\Big)^2}\Big\} \;,
\ee
where the upper limit is determined by
$A_0(\hat{z})=1-\Sigma$.
For a constant strong field, $A_0=Ez$, we recover the exponential we obtained in the previous section, $\eqref{general-z-int}=\eqref{expWithoutFourier2}$.

Let us for simplicity consider $N=2$, and then $2\Sigma=\omega_1=-\omega_2$. For Sauter-like fields with Fourier transform of the form~\eqref{exponential-Fourier} the total exponential is then given by
\be
P_2\sim\exp-4\Big\{\frac{\Sigma}{\omega_*}+\int_0^{\hat{z}}\!\ud z\sqrt{1-(A_0+\Sigma)^2}\Big\} \;,
\ee    
where $\Sigma$ is determined from its saddle point equation
\be
\frac{1}{\omega_*}=\int_0^{\hat{z}}\!\ud z\frac{A_0+\Sigma}{\sqrt{1-(A_0+\Sigma)^2}} \;.
\ee
Upon identifying $\Sigma$ with the constant $b$ in~\cite{Schneider:2014mla,SchneiderThesis}, we have thus recovered the results therein. (For a Sauter pulse we have $\omega_*=2\omega/\pi$.)
From this result we conclude that, for these Sauter-like fields, the exponential obtained by treating both the strong, spatially dependent field and the weak, time dependent fields nonperturbatively as in~\cite{Schneider:2014mla} can also be obtained by treating the weak field perturbatively to second order.

\end{document}